\pdfoutput=1
\documentclass[11pt]{article}
\usepackage[margin=1in]{geometry}
\usepackage{amsmath,amssymb,enumitem,ulem}
\usepackage{graphicx}
\usepackage[colorlinks=true,linkcolor=blue,citecolor=blue,urlcolor=blue]{hyperref}
\usepackage[numbers,sort&compress]{natbib}

\begin{document}

\title{Thermodynamic, Optical, and Orbital Signatures of Regular Asymptotically Flat Black Holes in Quasi-Topological Gravity}
\author{Zainab Malik,\\ \textit{Institute of Applied Sciences and Intelligent Systems, H-15 Islamabad, Pakistan}\\\texttt{zainabmalik8115@outlook.com}}
\date{\today}
\maketitle

\begin{abstract}
This study provides an analytic and numerical characterization of a class of regular, asymptotically flat black holes described by a deformed static spherical metric. The model is grounded in a four-dimensional non-polynomial quasi-topological framework in which higher-curvature corrections remain dynamically nontrivial while the static spherical sector retains a reduced-order structure, enabling tractable black-hole solutions with regular cores. Starting from the existence conditions of horizons and regularity, the allowed parameter domain and the extremal bound are derived. Hawking temperature, shadow radius, photon-ring Lyapunov exponent, and ISCO binding efficiency are then analyzed across the physically allowed parameter space. We further extend the analysis to Novikov--Thorne thin-disk accretion by deriving the flux kernel, effective-temperature profile, and bolometric luminosity scaling, and by providing representative numerical datasets for these quantities. A coherent trend emerges: increasing the deformation parameter drives the solution away from Schwarzschild behavior, reducing temperature, shadow size, and photon-orbit instability rate while enhancing orbital binding efficiency and accretion luminosity; increasing the exponent $\nu$ suppresses deformation effects and restores Schwarzschild-like observables. These results provide a compact phenomenological map linking horizon structure, thermodynamics, optical signatures, dynamical instability, and thin-disk accretion diagnostics in this regular black-hole family.
\end{abstract}

\section{Introduction}

Regular black holes are valuable laboratories for testing how short-distance modifications of gravity can resolve central singularities while preserving classical black-hole phenomenology at large radii. One of the approaches is related to adding a higher curvature correction to the Einstein action \cite{Lovelock:1972vz,Boulware:1985wk,Wiltshire:1988uq,Wheeler:1985nh,Glavan:2019inb,Konoplya:2020ibi,Aoki:2020iwm}. In particular, an infinite series of higher curvature corrections was used in quasi-topological gravity \cite{Bueno:2024dgm,Frolov:2024hhe,Bueno:2024eig,Konoplya:2024kih,Bueno:2025tli}. While those theories introduced non-trivial corrections only in theories higher than four, non-polynomial quasi-topological gravity (NP-QTG) \cite{Bueno:2025zaj,Borissova:2026wmn,Dubinsky:2026wcv,Lutfuoglu:2026gis} leads to corrections in four-dimensional spacetime.

Before introducing the metric ansatz, it is useful to summarize the essential structure of non-polynomial quasi-topological gravity (NP-QTG). The theory extends Einstein gravity with carefully organized higher-curvature contributions that remain dynamically nontrivial in four dimensions while avoiding the generic higher-derivative structure expected in arbitrary curvature expansions. In particular, whereas purely polynomial quasi-topological densities in four dimensions are typically topological or dynamically degenerate, non-polynomial completions can yield genuinely new dynamics in the static spherical sector.

A convenient schematic form of the action is \cite{Bueno:2025zaj,Borissova:2026wmn} 
\begin{equation}
S=\frac{1}{16\pi G}\int d^4x\,\sqrt{-g}\,\Big[R+\mathcal{F}(\mathcal{R})\Big],
\label{eq:npqtg-action}
\end{equation}
where $\mathcal{R}$ denotes a specific higher-curvature scalar built from Riemann-tensor contractions and $\mathcal{F}$ is chosen so that the physically relevant branch around maximally symmetric backgrounds does not introduce additional ghost-like tensor excitations. Under static spherical symmetry, the field equations are reduced to a lower-order master relation for the metric function (equivalently, a total-derivative structure in a suitable combination of equations), which is precisely what makes analytic regular black-hole families tractable.

Within this framework, the asymptotically flat metric function analyzed in Ref.~\cite{Konoplya:2026gim,Tsuda:2026xjc} is considered,
\begin{equation}\label{metric}
  f(r)=1-\frac{2Mr^{\mu-1}}{\left(r^{\nu}+\alpha^{\nu}/(2M)^{\nu/3}\right)^{\mu/\nu}}.
\end{equation}

The model is controlled by three effective parameters: a deformation scale $\alpha$ (or equivalently $\beta$), a regularity exponent $\mu$, and an additional exponent $\nu$ that governs how rapidly the metric interpolates between near-core regular behavior and asymptotic Schwarzschild form. From an observational perspective, a viable regular black-hole model should satisfy four simultaneous requirements: 
\begin{enumerate}[label=(\roman*),itemsep=0pt]
    \item geometric consistency (regular center and asymptotic flatness),
    \item a proper event-horizon sector,
    \item realistic thermodynamics, and
    \item measurable optical and orbital signatures.
\end{enumerate}
This analysis maps these requirements onto the same parameter space $(\mu,\nu,\beta)$. This unified perspective is important because each observable probes a different region of the effective potential: Hawking temperature is tied to near-horizon gradients, the shadow and Lyapunov exponent are controlled by unstable null circular orbits, and ISCO efficiency is determined by stable timelike dynamics. Interpreting all four together provides a robust phenomenological characterization of the same underlying metric deformation.

\section{Regular asymptotically flat black holes: parameter domain}

Static, spherically symmetric geometries are considered:
\begin{equation}\label{lineelement}
  ds^2=-f(r)dt^2+\frac{dr^2}{f(r)}+r^2d\Omega^2,
\end{equation}
with metric function is given by eq. \ref{metric} (see \cite{Konoplya:2026gim,Tsuda:2026xjc}).
We define
\begin{equation}
  g\equiv \frac{\alpha}{(2M)^{1/3}},
  \qquad
  x\equiv\frac{r}{2M},
  \qquad
  \beta\equiv\frac{g}{2M}=\frac{\alpha}{(2M)^{4/3}},
\end{equation}
so that
\begin{equation}
  f(x)=1-\frac{x^{\mu-1}}{\left(x^{\nu}+\beta^{\nu}\right)^{\mu/\nu}}.
\end{equation}
For $r\to\infty$,
\begin{equation}
  f(r)=1-\frac{2M}{r}+\mathcal{O}\!\left(r^{-1-\nu}\right),
\end{equation}
therefore the spacetime is asymptotically flat for $\nu>0$.

Near the center,
\begin{equation}
  f(r)=1-\frac{2M}{g^{\mu}}r^{\mu-1}+\mathcal{O}\!\left(r^{\mu+\nu-1}\right),
\end{equation}
which implies $m(r)=\frac{r}{2}[1-f(r)]\sim r^{\mu}$. Finiteness of curvature invariants requires $m(r)=\mathcal{O}(r^3)$, hence
\begin{equation}
  \mu\ge 3.
\end{equation}

Horizons are given by $f(x_h)=0$:
\begin{equation}
  x^{\mu-1}=\left(x^{\nu}+\beta^{\nu}\right)^{\mu/\nu}
  \quad\Longleftrightarrow\quad
  \beta^{\nu}=x^{\nu(\mu-1)/\mu}-x^{\nu}.
\end{equation}
It is convenient to introduce the function
\begin{equation}
  h(x)=x^{\nu(\mu-1)/\mu}-x^{\nu},
\end{equation}
its maximum occurs at
\begin{equation}
  x_\ast=\left(\frac{\mu-1}{\mu}\right)^{\mu/\nu},
  \qquad
  h_{\max}=\frac{1}{\mu}\left(\frac{\mu-1}{\mu}\right)^{\mu-1},
\end{equation}
so the critical deformation is
\begin{equation}
  \beta_c=\left[\frac{1}{\mu}\left(\frac{\mu-1}{\mu}\right)^{\mu-1}\right]^{1/\nu}.
\end{equation}
Thus
\begin{equation}
  \begin{cases}
    0<\beta<\beta_c, & \text{two horizons }(r_-,r_+)\ \text{(nonextremal BH)},\\
    \beta=\beta_c, & \text{one degenerate horizon (extremal BH)},\\
    \beta>\beta_c, & \text{no horizon}.
  \end{cases}
\end{equation}
Thus, we conclude that the regular asymptotically flat black-hole domain is
\begin{equation}
  \nu>0,\qquad \mu\ge 3,\qquad 0<\beta\le\beta_c,
\end{equation}
or equivalently
\begin{equation}
  0<\frac{\alpha}{(2M)^{4/3}}\le
  \left[\frac{1}{\mu}\left(\frac{\mu-1}{\mu}\right)^{\mu-1}\right]^{1/\nu}.
\end{equation}

This domain separates physically acceptable black holes from horizonless compact objects within the same metric family. The condition $\mu\ge 3$ guarantees central regularity, while $\beta\le\beta_c$ guarantees horizon formation. Therefore, $\beta_c$ acts as the organizing scale of the whole phenomenology: moving toward $\beta_c$ corresponds to approaching extremality, where thermodynamic and optical observables are expected to deviate most strongly from Schwarzschild values.

\section{Hawking temperature: analytic form and parameter dependence}

For the line element above, the Hawking temperature is
\begin{equation}
  T_H=\frac{\kappa}{2\pi}=\frac{f'(r_+)}{4\pi},
\end{equation}
where $\kappa$ is the surface gravity, $r_+$ is the outer horizon and
\begin{equation}
  f'(r)=\frac{2M\,r^{\mu-2}\left[r^{\nu}-(\mu-1)g^{\nu}\right]}{\left(r^{\nu}+g^{\nu}\right)^{1+\mu/\nu}}.
\end{equation}
Using the horizon equation
\begin{equation}
  2M\,r_+^{\mu-1}=\left(r_+^{\nu}+g^{\nu}\right)^{\mu/\nu},
\end{equation}
one obtains the compact analytic result
\begin{equation}
  T_H=\frac{1}{4\pi}\frac{r_+^{\nu}-(\mu-1)g^{\nu}}{r_+\left(r_+^{\nu}+g^{\nu}\right)}.
  \label{eq:TH-physical}
\end{equation}
In dimensionless form ($x_+=r_+/(2M)$),
\begin{equation}
  \tilde T\equiv 8\pi M T_H
  =\frac{x_+^{\nu}-(\mu-1)\beta^{\nu}}{x_+\left(x_+^{\nu}+\beta^{\nu}\right)}.
  \label{eq:TH-dimless}
\end{equation}
Equation~\eqref{eq:TH-dimless} shows that $\tilde T\to1$ in the Schwarzschild limit ($\beta\to0$, $x_+\to1$), while $\tilde T=0$ at extremality, where $x_+^{\nu}=(\mu-1)\beta^{\nu}$.

For the parameter scans below, $f(x)=0$ is solved numerically for the outer horizon $x_+$, and Eq.~\eqref{eq:TH-dimless} is then evaluated.

\begin{figure}[t]
  \centering
  \includegraphics[width=0.78\linewidth]{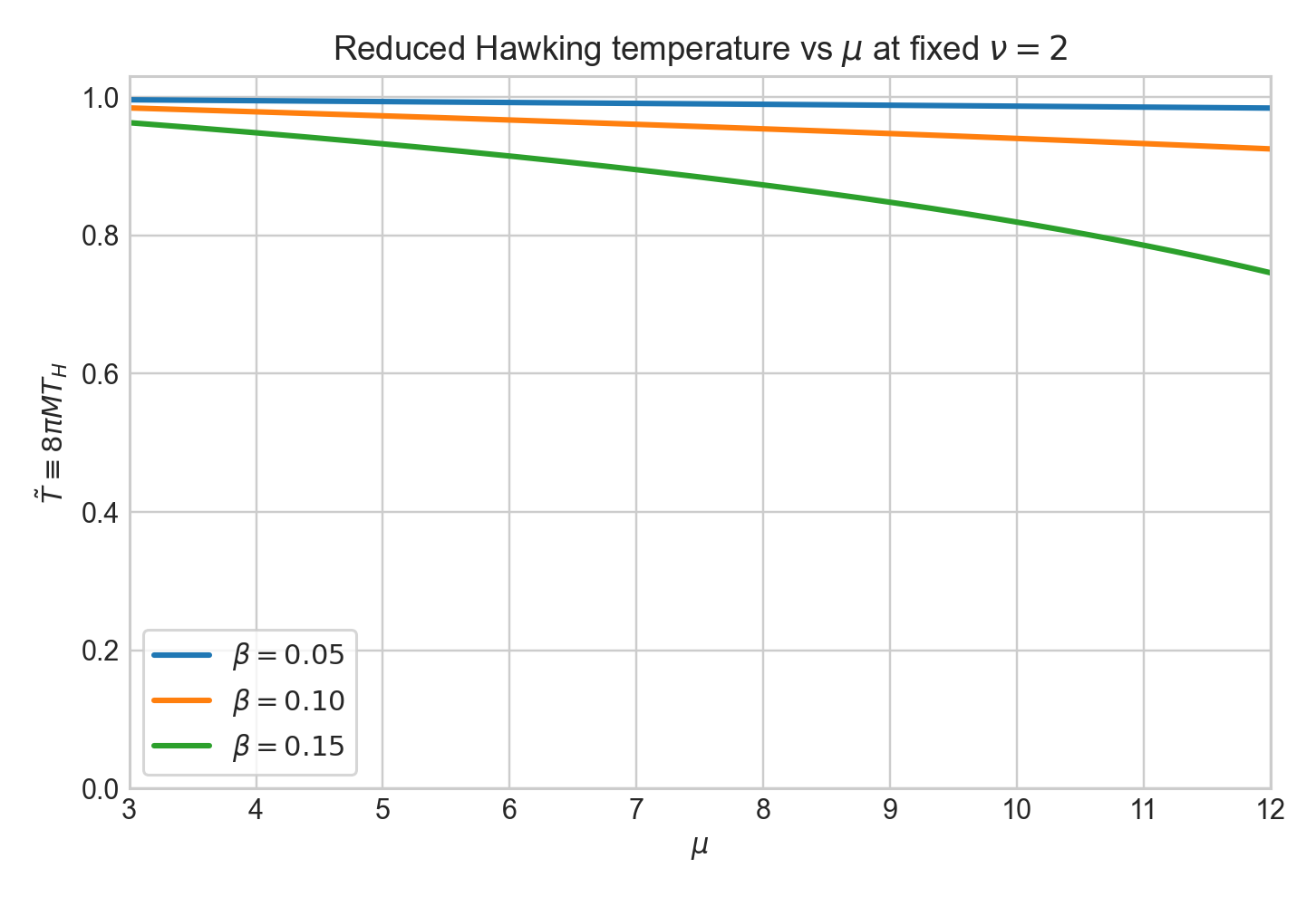}
  \caption{Reduced Hawking temperature $\tilde T=8\pi M T_H$ versus $\mu$ at fixed $\nu=2$, for selected $\beta$. Increasing $\mu$ lowers $\tilde T$, and the suppression is stronger for larger deformation parameter $\beta$.}
  \label{fig:temp-vs-mu}
\end{figure}

\begin{figure}[t]
  \centering
  \includegraphics[width=0.78\linewidth]{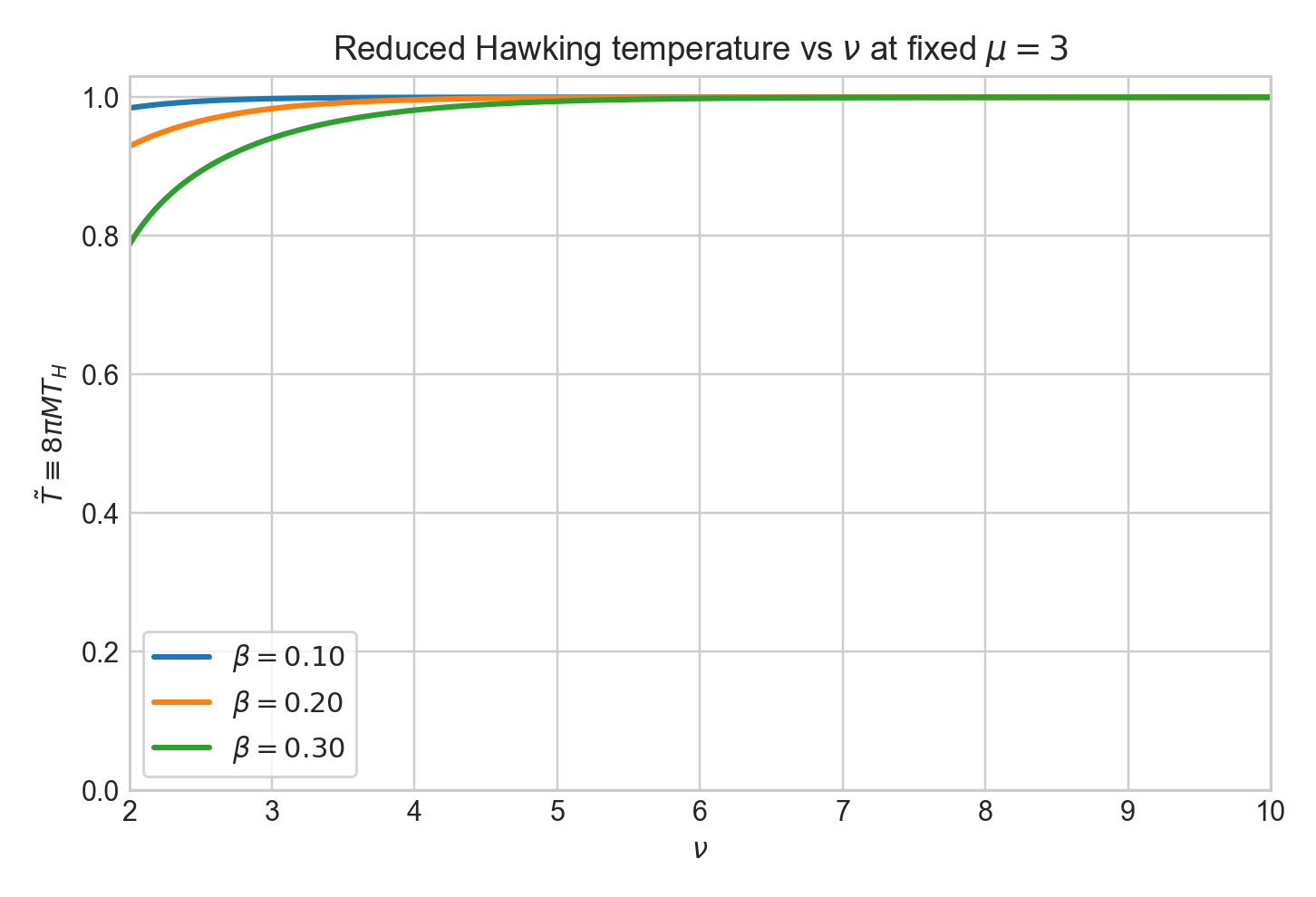}
  \caption{Reduced Hawking temperature $\tilde T=8\pi M T_H$ versus $\nu$ at fixed $\mu=3$, for selected $\beta$. For fixed $\mu$ and $\beta$, $\tilde T$ increases with $\nu$ and approaches the Schwarzschild value $\tilde T\to1$.}
  \label{fig:temp-vs-nu}
\end{figure}

These temperature trends are consistent with the existence domain from the previous section. For larger $\beta$, the outer horizon moves toward the extremal regime and the surface gravity decreases, so $T_H$ is reduced and eventually vanishes at extremality. By contrast, increasing $\nu$ at fixed $(\mu,\beta)$ suppresses the relative weight of the deformation term $\beta^{\nu}$ and restores Schwarzschild-like thermodynamics (see figs. \ref{fig:temp-vs-mu}, \ref{fig:temp-vs-nu}). In this sense, $\beta$ controls the strength of regular-core effects, while $\nu$ controls how efficiently those effects decouple in horizon-scale physics.

\section{Shadow radius: analytic conditions and parameter dependence}

Black-hole shadows provide one of the most direct strong-field probes of spacetime geometry because their angular size and shape are determined by unstable null orbits near the photon sphere rather than by weak-field dynamics. This makes shadow observables complementary to thermodynamic or orbital diagnostics: they are sensitive to near-horizon metric deformations and can be confronted with horizon-scale imaging measurements. Therefore, a large number of recent publications has been devoted to calculations of black-hole shadows' parameters \cite{Perlick:2015vta,Zakharov:2025rhb,Bambi:2019tjh,Younsi:2016azx,Tsukamoto:2017fxq,Spina:2025wxb,Kumar:2018ple,Konoplya:2021slg,Konoplya:2019fpy,Afrin:2021imp,Perlick:2021aok,Bambi:2015kza,Hioki:2009na,Lutfuoglu:2025ldc,Cunha:2018acu,Zakharov:2025cnq,Khodadi:2020jij,Lutfuoglu:2026fks}. In regular black-hole models, the shadow is especially informative because it encodes how singularity-resolving corrections propagate outward into the observable photon region.

For null geodesics in a static, spherically symmetric metric, the effective potential is $V_{\rm eff}=L^2 f(r)/r^2$. Circular photon orbits satisfy
\begin{equation}
  \frac{d}{dr}\left(\frac{f(r)}{r^2}\right)=0
  \quad\Longleftrightarrow\quad
  2f(r_{\rm ph})=r_{\rm ph}f'(r_{\rm ph}),
\end{equation}
where $r_{\rm ph}$ is the photon-sphere radius. For this metric, in dimensionless variables this condition can be written as
\begin{equation}
  x_{\rm ph}^{\mu-1}\left[3x_{\rm ph}^{\nu}+(3-\mu)\beta^{\nu}\right]
  -2\left(x_{\rm ph}^{\nu}+\beta^{\nu}\right)^{1+\mu/\nu}=0,
  \label{eq:photon-condition}
\end{equation}
where $x_{\rm ph}=r_{\rm ph}/2M$.

The observable shadow radius for a static observer at infinity is the critical impact parameter,
\begin{equation}
  R_{\rm sh}=\frac{r_{\rm ph}}{\sqrt{f(r_{\rm ph})}},
  \qquad
  \tilde R_{\rm sh}\equiv\frac{R_{\rm sh}}{2M}=\frac{x_{\rm ph}}{\sqrt{f(x_{\rm ph})}}.
  \label{eq:shadow-radius}
\end{equation}
In the Schwarzschild limit ($\beta\to0$), one recovers $\tilde R_{\rm sh}=3\sqrt{3}/2\simeq2.598$.

Numerically, for each parameter set, $f(x_+)=0$ is solved for the outer horizon $x_+$, Eq.~\eqref{eq:photon-condition} is then solved for the outer photon sphere $x_{\rm ph}>x_+$, and Eq.~\eqref{eq:shadow-radius} is finally evaluated.

\begin{figure}[t]
  \centering
  \includegraphics[width=0.78\linewidth]{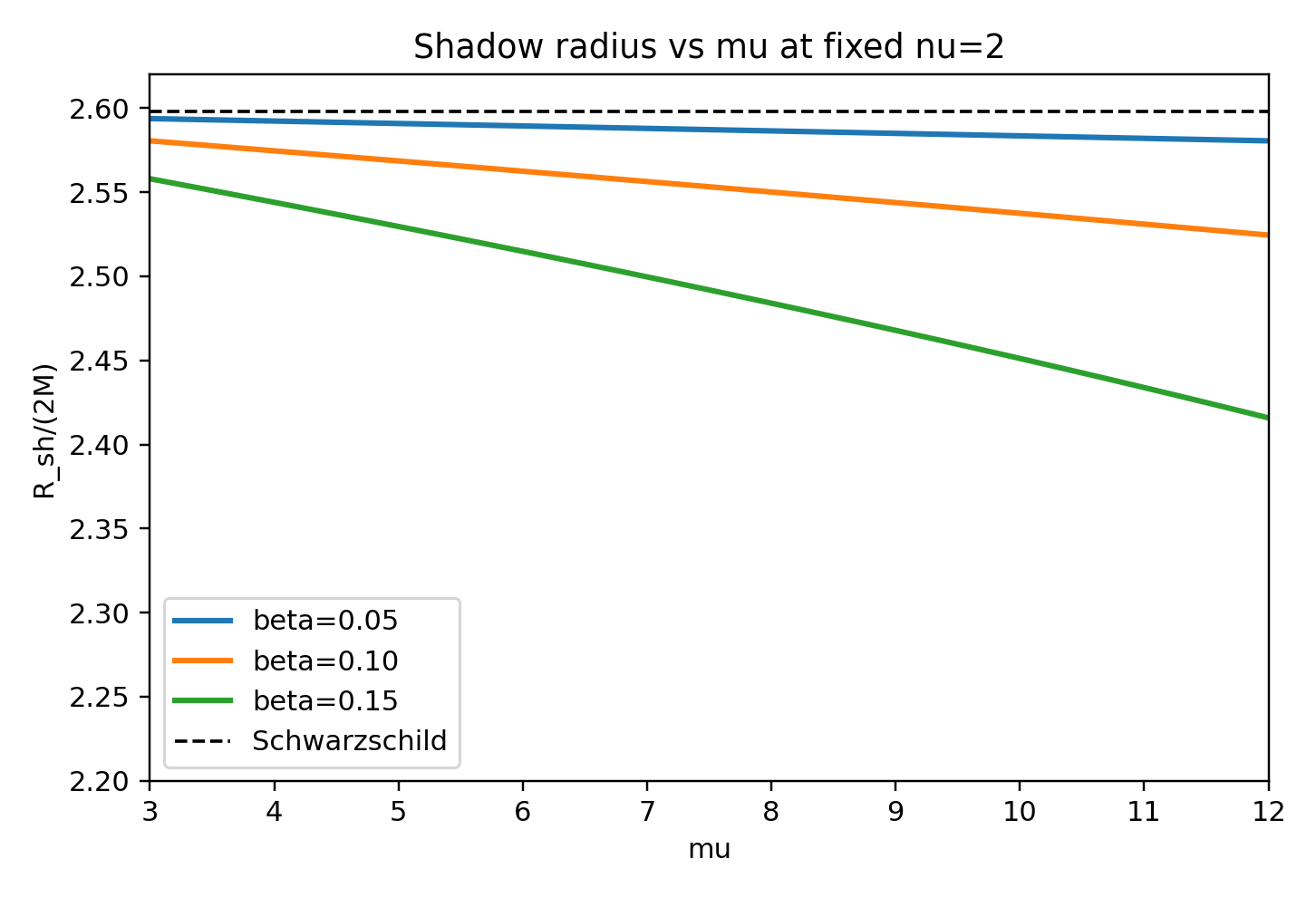}
  \caption{Reduced shadow radius $\tilde R_{\rm sh}=R_{\rm sh}/(2M)$ versus $\mu$ at fixed $\nu=2$, for selected $\beta$. Larger $\mu$ decreases the shadow size, with a stronger effect at larger $\beta$.}
  \label{fig:shadow-vs-mu}
\end{figure}

\begin{figure}[t]
  \centering
  \includegraphics[width=0.78\linewidth]{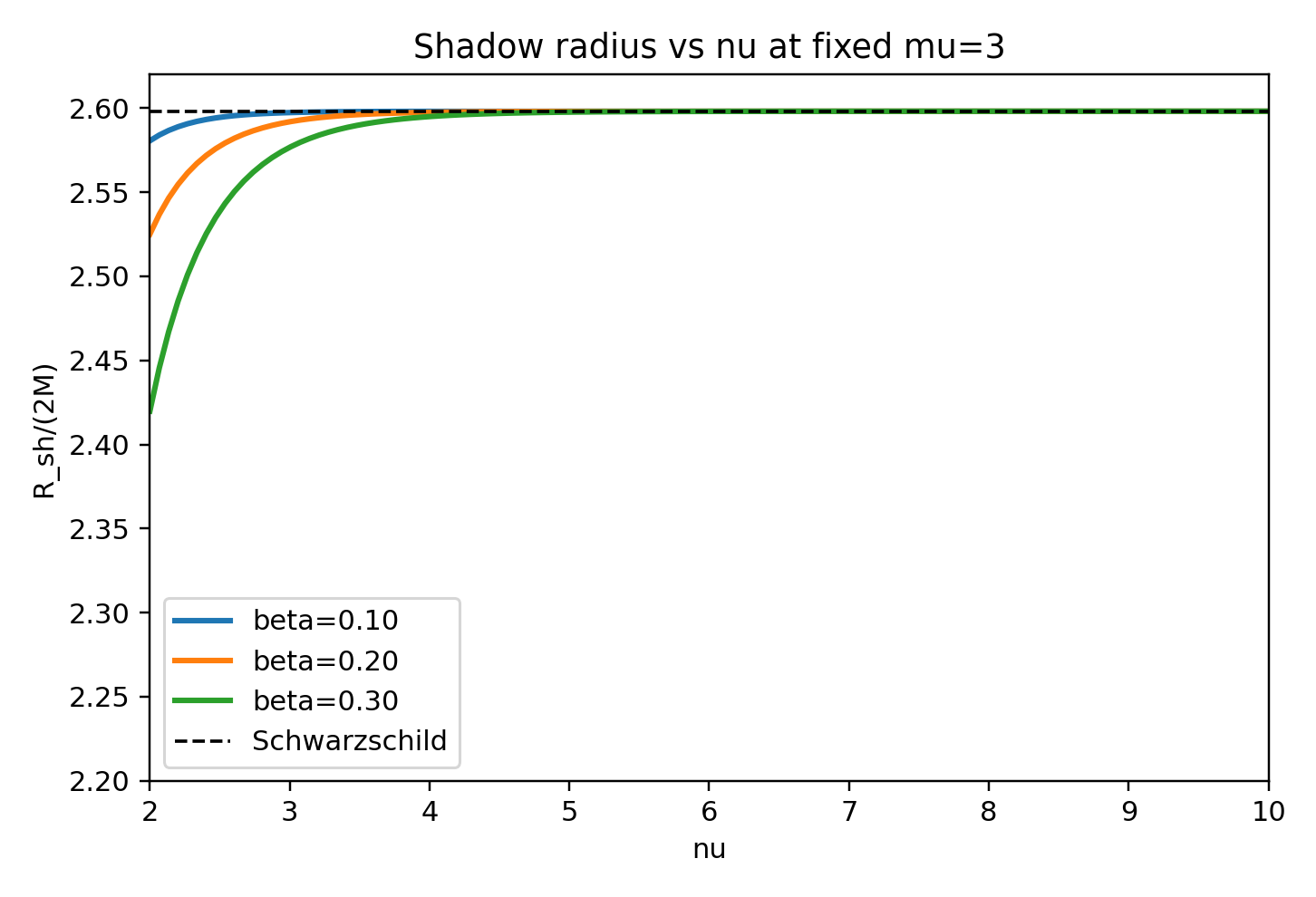}
  \caption{Reduced shadow radius $\tilde R_{\rm sh}$ versus $\nu$ at fixed $\mu=3$, for selected $\beta$. Increasing $\nu$ restores the Schwarzschild value from below.}
  \label{fig:shadow-vs-nu}
\end{figure}

\begin{figure}[t]
  \centering
  \includegraphics[width=0.78\linewidth]{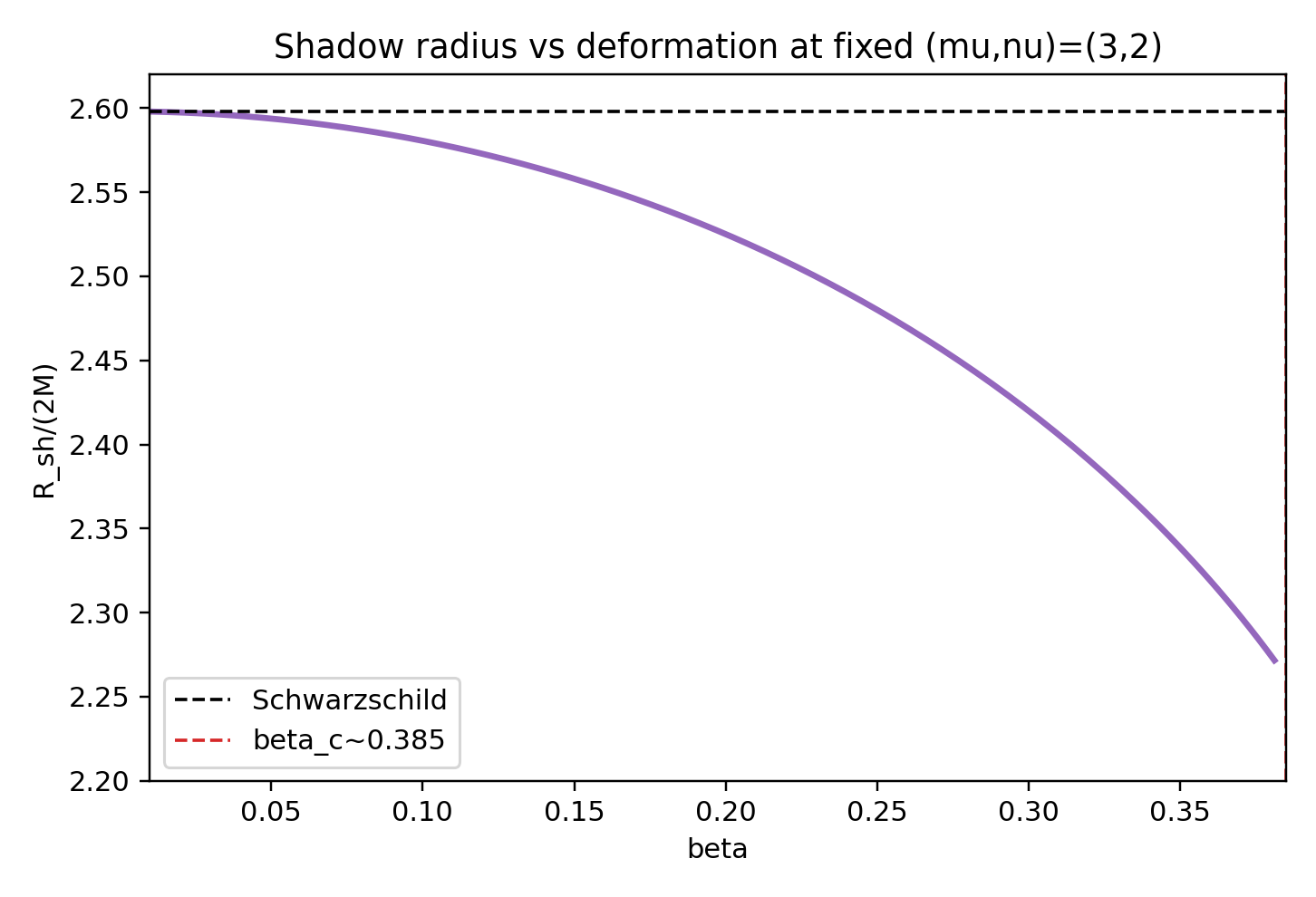}
  \caption{Reduced shadow radius $\tilde R_{\rm sh}$ versus deformation parameter $\beta$ at fixed $(\mu,\nu)=(3,2)$. The dashed vertical line marks the extremal bound $\beta_c\approx0.385$; approaching extremality reduces the shadow radius.
  }
  \label{fig:shadow-vs-beta}
\end{figure}

The shadow sector confirms the same hierarchy observed in thermodynamics. Stronger deformation (larger $\beta$) shifts the unstable photon orbit inward, shrinking both $r_{\rm ph}$ and the impact-parameter radius $R_{\rm sh}$. Increasing $\nu$ at fixed $(\mu,\beta)$ has the opposite effect, pushing the geometry toward Schwarzschild and raising $R_{\rm sh}$ toward $3\sqrt{3}M$. The monotonic behavior across all scans indicates that shadow observables can serve as clean diagnostics of how close the solution is to the extremal boundary.

\section{Photon-sphere Lyapunov exponent and instability timescale}

The instability rate of the outer photon sphere is measured by the Lyapunov exponent with respect to coordinate time. For static spherical metrics, it can be written as
\begin{equation}
  \lambda=\sqrt{\frac{f(r_{\rm ph})\left[2f(r_{\rm ph})-r_{\rm ph}^{2}f''(r_{\rm ph})\right]}{2r_{\rm ph}^{2}}}.
  \label{eq:lyapunov-physical}
\end{equation}
In dimensionless variables ($x_{\rm ph}=r_{\rm ph}/2M$), define
\begin{equation}
  \tilde\lambda\equiv 2M\lambda
  =\sqrt{\frac{f(x_{\rm ph})\left[2f(x_{\rm ph})-x_{\rm ph}^{2}f''(x_{\rm ph})\right]}{2x_{\rm ph}^{2}}}.
  \label{eq:lyapunov-dimless}
\end{equation}
In the Schwarzschild limit ($\beta\to0$), one obtains
\begin{equation}
  \tilde\lambda_{\rm Schw}=\frac{2}{3\sqrt{3}}\simeq0.3849.
\end{equation}

\begin{figure}[t]
  \centering
  \includegraphics[width=0.78\linewidth]{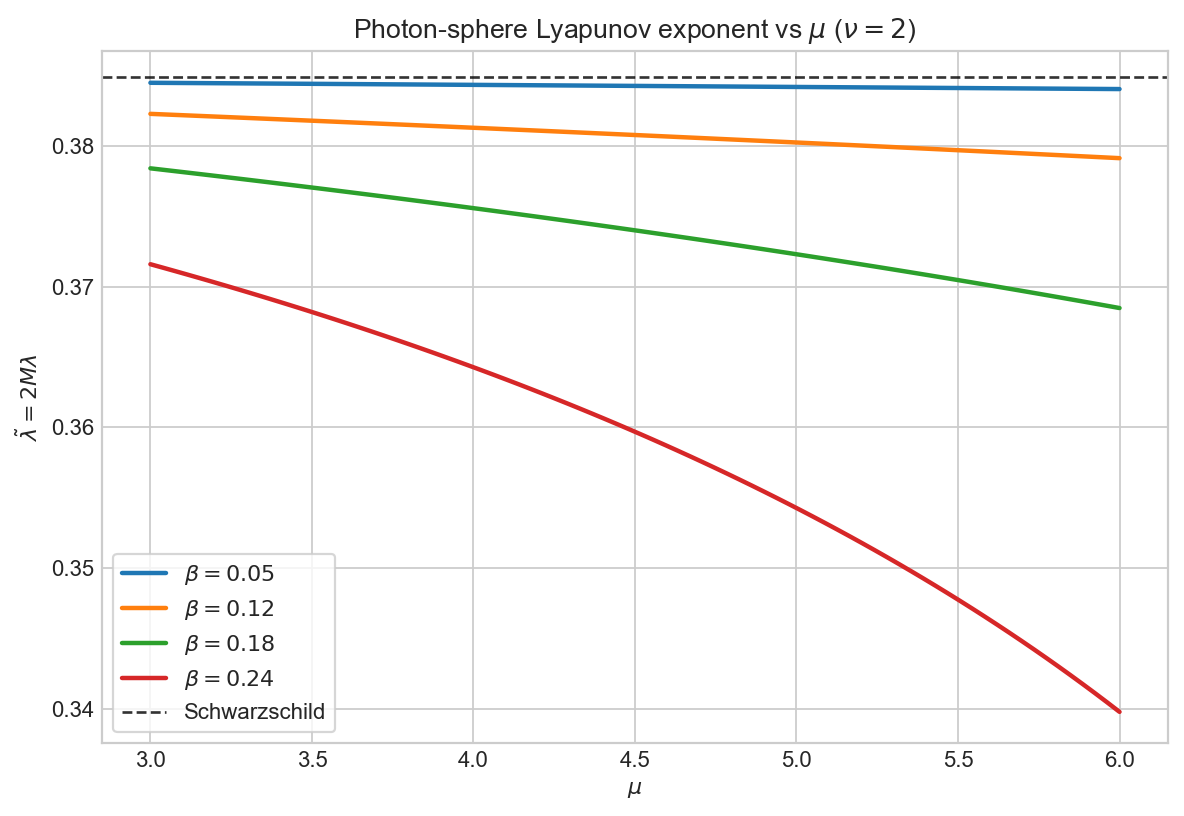}
  \caption{Reduced Lyapunov exponent $\tilde\lambda=2M\lambda$ versus $\mu$ at fixed $\nu=2$, for selected $\beta$. Increasing $\mu$ reduces the photon-sphere instability rate, with stronger suppression at larger $\beta$.}
  \label{fig:lyapunov-vs-mu}
\end{figure}

\begin{figure}[t]
  \centering
  \includegraphics[width=0.78\linewidth]{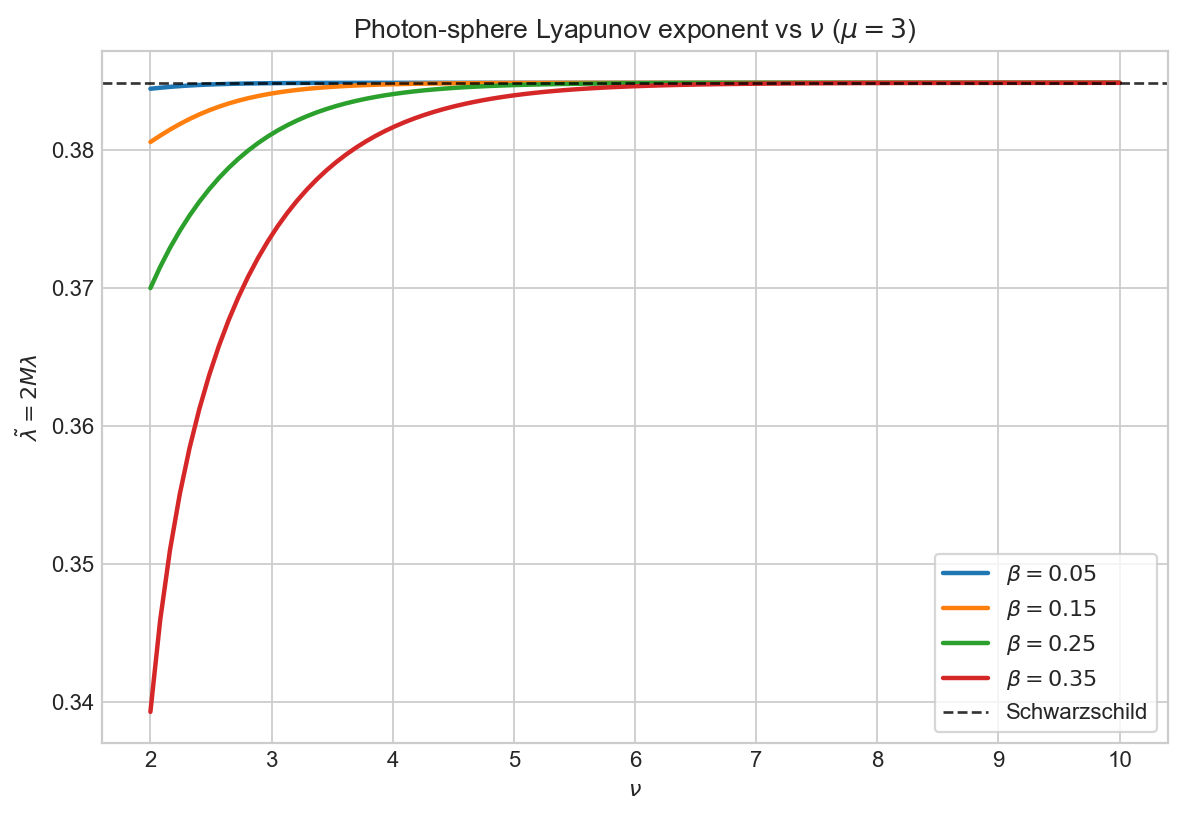}
  \caption{Reduced Lyapunov exponent $\tilde\lambda$ versus $\nu$ at fixed $\mu=3$, for selected $\beta$. Increasing $\nu$ drives $\tilde\lambda$ toward the Schwarzschild value $2/(3\sqrt{3})$.}
  \label{fig:lyapunov-vs-nu}
\end{figure}

\begin{figure}[t]
  \centering
  \includegraphics[width=0.78\linewidth]{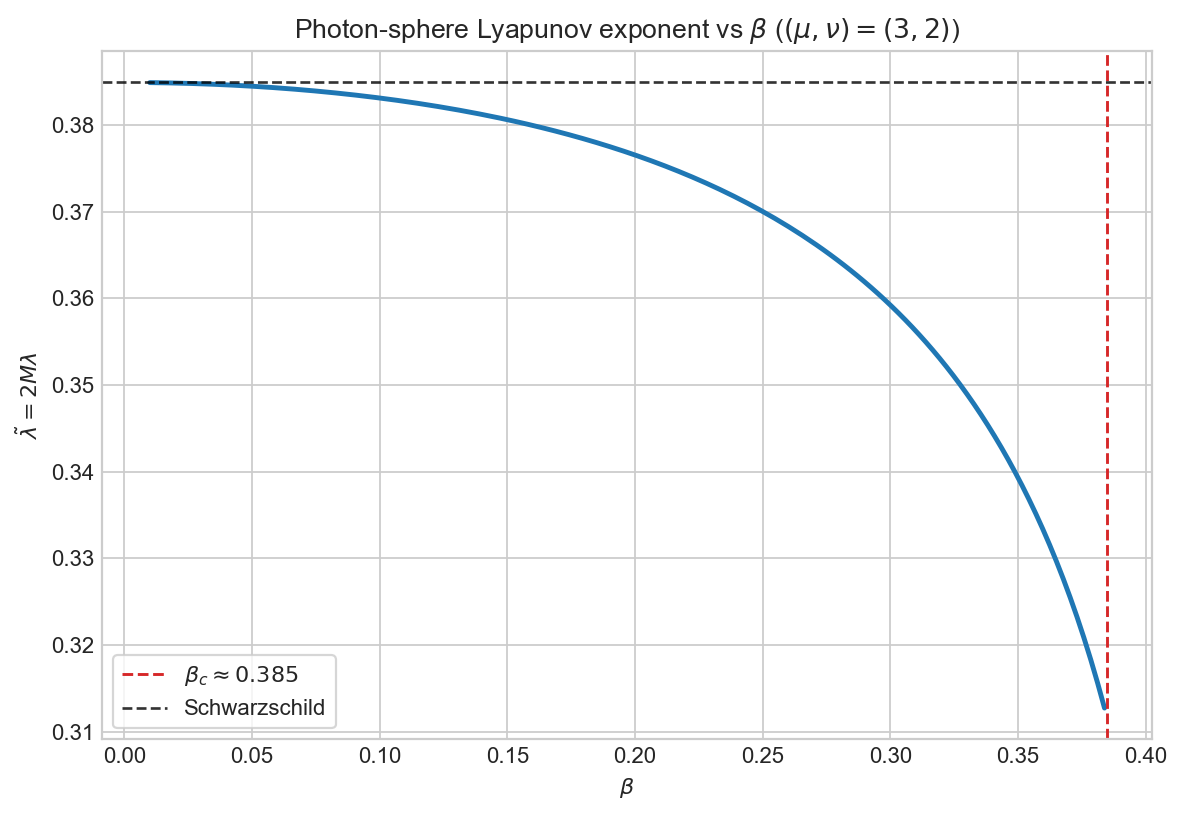}
  \caption{Reduced Lyapunov exponent $\tilde\lambda$ versus deformation parameter $\beta$ at fixed $(\mu,\nu)=(3,2)$. The dashed vertical line marks the extremal bound $\beta_c\approx0.385$; increasing $\beta$ decreases the instability rate.}
  \label{fig:lyapunov-vs-beta}
\end{figure}

The Lyapunov sector is complementary to the shadow analysis because it measures not only the location of the photon sphere but also the sharpness of the local potential barrier. Stronger deformation lowers $\tilde\lambda$, implying longer instability timescales for null orbits and weaker eikonal damping rates. Increasing $\nu$ reverses this trend and restores Schwarzschild-like instability. In the eikonal limit, this has a direct correspondence with quasinormal modes, where the real part is set by the photon-orbit angular frequency and the imaginary part by the Lyapunov instability rate:
\begin{equation}
  \omega_{\ell n}\simeq \ell\,\Omega_{\rm ph}-i\left(n+\tfrac{1}{2}\right)\lambda.
\end{equation}
Consequently, the ringdown damping time is inversely controlled by the same instability scale,
\begin{equation}
  \tau_{\ell n}=\frac{1}{|\operatorname{Im}\omega_{\ell n}|}\simeq\frac{1}{\left(n+\tfrac{1}{2}\right)\lambda},
\end{equation}
so a smaller $\lambda$ near extremality implies longer-lived eikonal modes. Notice, that the correspondence may not hold for arbitrary theory of gravity, and there are exceptions discussed in  \cite{Khanna:2016yow,Konoplya:2022gjp,Konoplya:2017wot,Bolokhov:2023dxq}.
Therefore, the observed suppression of $\lambda$ toward extremality is consistent with more weakly damped ringdown modes in the same parameter region. Together with $R_{\rm sh}$, this provides a two-parameter optical--dynamical diagnostic of how close the solution is to extremality.

\section{Binding energy of massive particles and ISCO efficiency}

Among geodesic observables, the ISCO binding efficiency is one of the simplest and most directly interpretable characteristics of a black-hole spacetime. It gives a first-principles proxy for accretion efficiency, quantifying the maximal fraction of rest-mass energy that can be released before matter plunges from the inner edge of a thin disk~\cite{NovikovThorne1973}.

For equatorial timelike geodesics, the radial equation can be written as
\begin{equation}
  \dot r^2+V_{\rm eff}(r;L)=E^2,
  \qquad
  V_{\rm eff}(r;L)=f(r)\left(1+\frac{L^2}{r^2}\right),
\end{equation}
where $E$ and $L$ are the specific energy and angular momentum.

For circular orbits ($\dot r=0$ and $dV_{\rm eff}/dr=0$), one obtains
\begin{equation}
  E^2=\frac{2f(r)^2}{2f(r)-rf'(r)},
  \qquad
  L^2=\frac{r^3 f'(r)}{2f(r)-rf'(r)}.
  \label{eq:EL-circular}
\end{equation}
The innermost stable circular orbit (ISCO) is determined by the marginal stability condition
\begin{equation}
  r f(r) f''(r)+3f(r)f'(r)-2r f'(r)^2=0.
  \label{eq:isco-condition}
\end{equation}

The binding efficiency is defined as
\begin{equation}
  \eta\equiv 1-E_{\rm ISCO},
  \qquad
  E_{\rm ISCO}=E\Biggr|_{r=r_{\rm ISCO}}.
  \label{eq:binding-efficiency}
\end{equation}
For Schwarzschild, Eq.~\eqref{eq:binding-efficiency} gives
\begin{equation}
  \eta_{\rm Schw}=1-\sqrt{\frac{8}{9}}\simeq 0.05719.
\end{equation}

Again, for each parameter set $(\mu,\nu,\beta)$ in the black-hole domain, $f(x_+)=0$ is numerically solved for the outer horizon, Eq.~\eqref{eq:isco-condition} is then solved for the outer timelike ISCO radius $x_{\rm ISCO}>x_+$, and $\eta$ is finally evaluated from Eq.~\eqref{eq:binding-efficiency}.

\begin{figure}[t]
  \centering
  \includegraphics[width=0.78\linewidth]{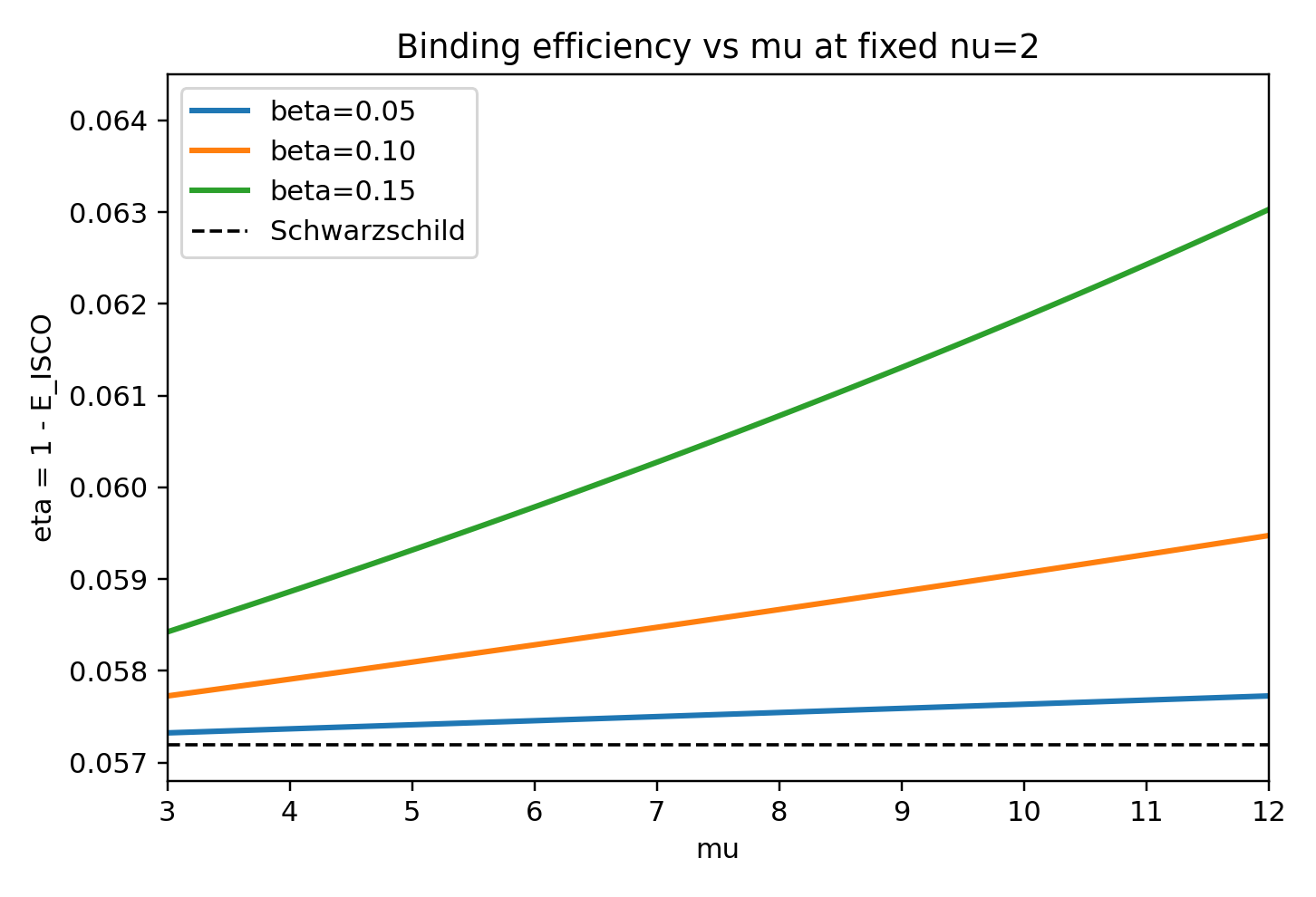}
  \caption{Binding efficiency $\eta=1-E_{\rm ISCO}$ versus $\mu$ at fixed $\nu=2$, for selected $\beta$. For fixed $\nu$, increasing $\mu$ increases $\eta$, with a stronger enhancement at larger $\beta$.}
  \label{fig:binding-vs-mu}
\end{figure}

\begin{figure}[t]
  \centering
  \includegraphics[width=0.78\linewidth]{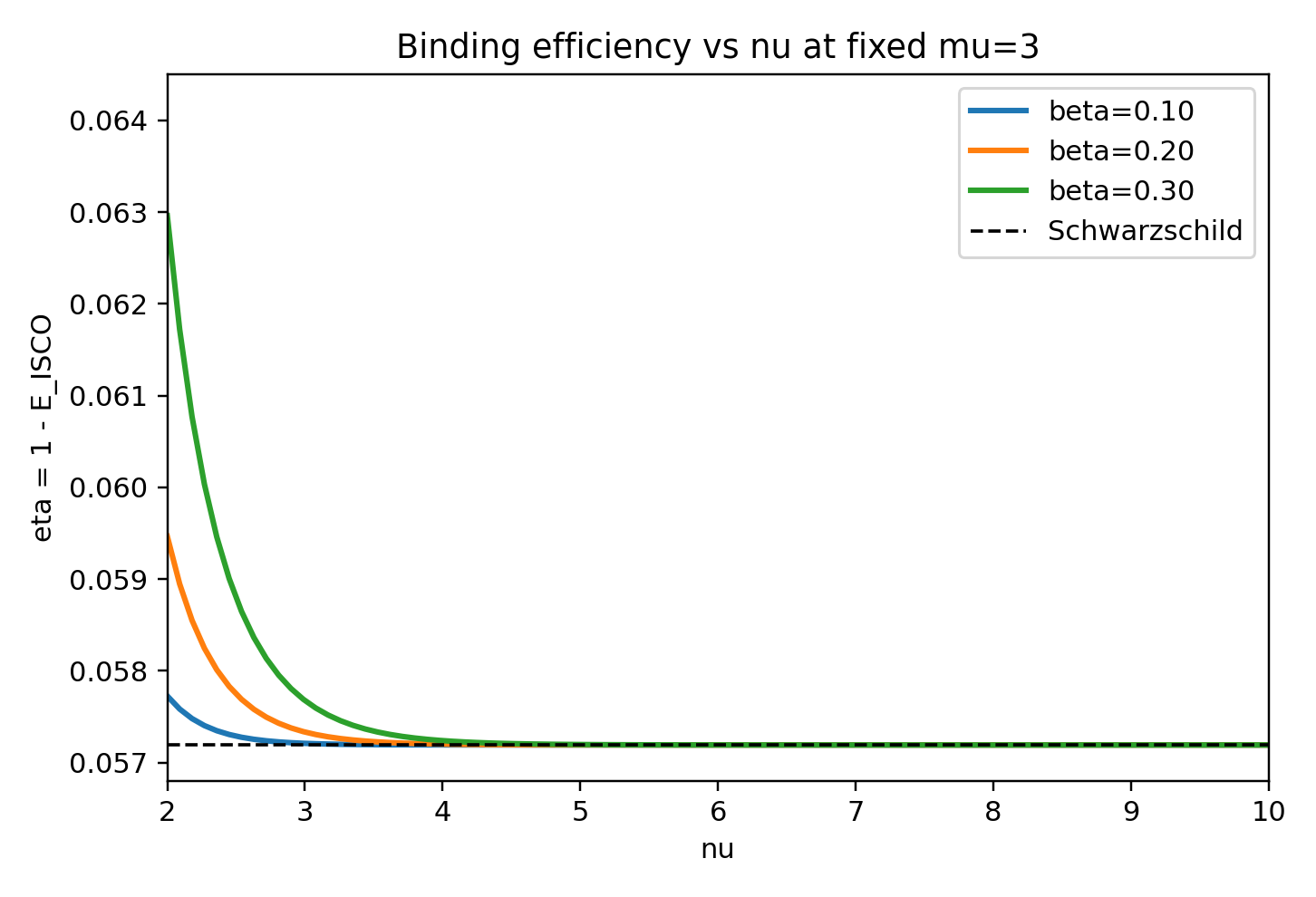}
  \caption{Binding efficiency $\eta$ versus $\nu$ at fixed $\mu=3$, for selected $\beta$. Increasing $\nu$ reduces deformation effects and drives $\eta$ toward the Schwarzschild value.}
  \label{fig:binding-vs-nu}
\end{figure}

\begin{figure}[t]
  \centering
  \includegraphics[width=0.78\linewidth]{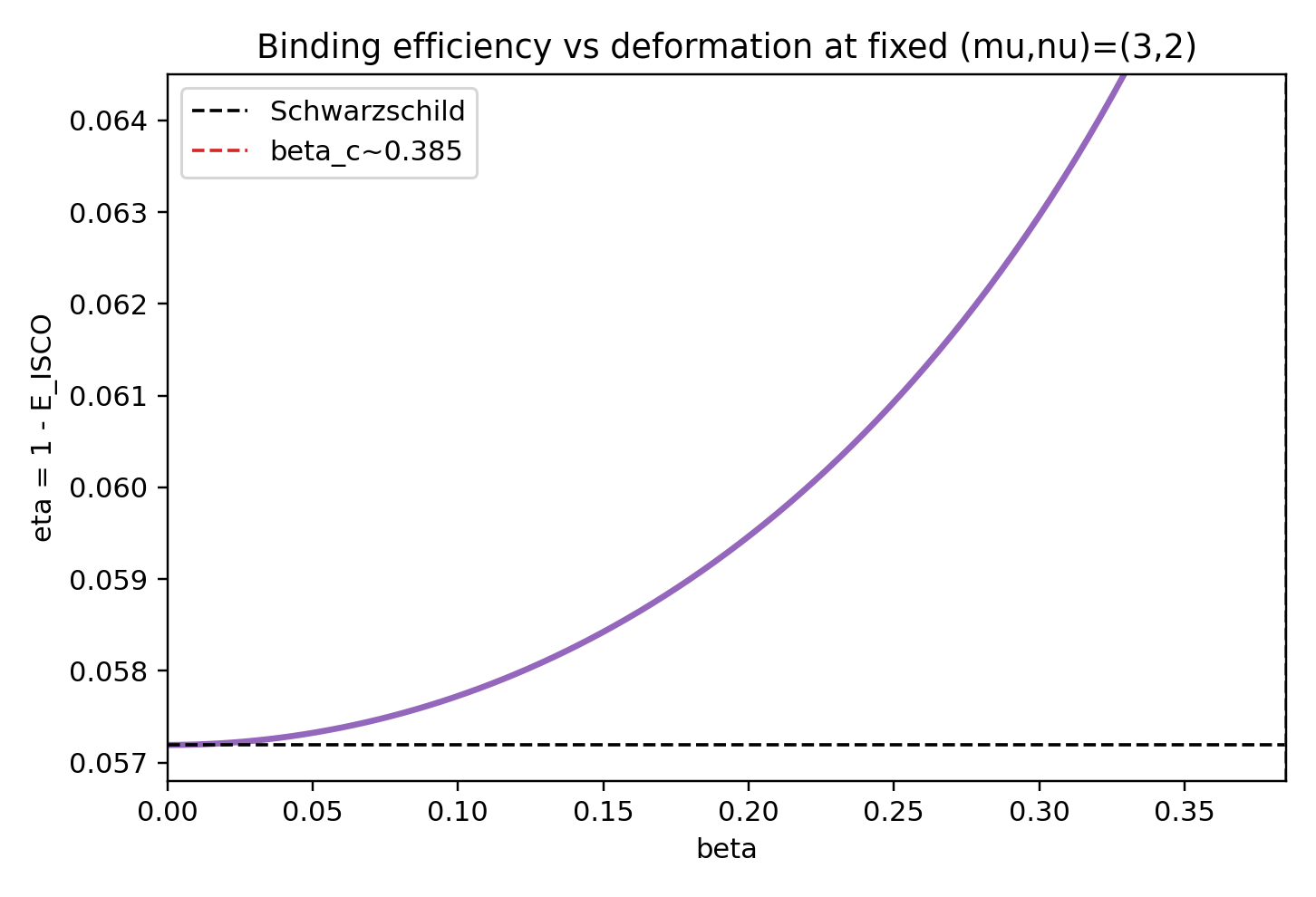}
  \caption{Binding efficiency $\eta$ versus deformation parameter $\beta$ at fixed $(\mu,\nu)=(3,2)$. As $\beta$ approaches the extremal bound $\beta_c$, $\eta$ increases above the Schwarzschild benchmark.}
  \label{fig:binding-vs-beta}
\end{figure}

Unlike temperature and shadow size, the ISCO efficiency grows with stronger deformation. Physically, the deformed geometry allows stable circular orbits to persist deeper in the potential well before the onset of instability, lowering $E_{\rm ISCO}$ and increasing $\eta=1-E_{\rm ISCO}$. Therefore, in this model, near-extremal configurations are simultaneously colder and optically smaller, yet more efficient at releasing orbital binding energy. Increasing $\nu$ reverses these effects by reducing deformation and returning the efficiency toward the Schwarzschild value.

\section{Thin-disk accretion process: flux, temperature, and luminosity}

To move from orbital diagnostics to an explicit accretion model, consider a stationary, geometrically thin, optically thick disk in the Novikov--Thorne framework~\cite{NovikovThorne1973}. The disk is assumed to lie in the equatorial plane, to radiate locally as a blackbody, and to have a stress-free inner edge at $r_{\rm ISCO}$.

For circular geodesic motion at the coordinate $r$ in a static spherical metric (\ref{lineelement}), the angular velocity is
\begin{equation}
  \Omega(r)\equiv\frac{d\phi}{dt}=\sqrt{\frac{f'(r)}{2r}}.
  \label{eq:omega-circular}
\end{equation}
Using Eq.~\eqref{eq:EL-circular}, the specific energy and angular momentum may be written as
\begin{equation}
  E(r)=\frac{\sqrt{2}\,f(r)}{\sqrt{2f(r)-rf'(r)}},
  \qquad
  L(r)=\sqrt{\frac{r^3f'(r)}{2f(r)-rf'(r)}}.
  \label{eq:EL-explicit}
\end{equation}

The radiative flux emitted from one face of the disk is
\begin{equation}
  F(r)=\frac{\dot M}{4\pi r}\,\mathcal{D}(r),
\end{equation}
with
\begin{equation}
  \mathcal{D}(r)=
  -\frac{d\Omega/dr}{\left[E(r)-\Omega(r) L(r)\right]^2}
  \intop_{r_{\rm ISCO}}^{r}\left[E(r')-\Omega(r') L(r')\right]\frac{dL}{dr'}\,dr'.
  \label{eq:disk-flux-shape}
\end{equation}
Here $\dot M$ is the mass accretion rate and all orbital quantities are evaluated on circular geodesics of the same metric. The effective temperature profile then follows from local blackbody emission,
\begin{equation}
  T_{\rm eff}(r)=\left(\frac{F(r)}{\sigma_{\rm SB}}\right)^{1/4}.
  \label{eq:teff-profile}
\end{equation}

In dimensionless variables ($x\equiv r/2M$), the flux takes the scaling form
\begin{equation}
  F(r)=\frac{\dot M}{16\pi M^2}\,\mathcal{F}(x;\mu,\nu,\beta),
\end{equation}
where
\begin{equation}
  \mathcal{F}(x;\mu,\nu,\beta)
  \equiv
  \frac{\mathcal{D}(x;\mu,\nu,\beta)}{x}.
  \label{eq:dimensionless-flux}
\end{equation}
Therefore, at fixed dimensionless parameters, $F\propto \dot M/M^2$ and $T_{\rm eff}\propto(\dot M/M^2)^{1/4}$, while $(\mu,\nu,\beta)$ determine the shape function $\mathcal{F}(x)$ through $f$, $f'$, and the location of $x_{\rm ISCO}$.

The bolometric luminosity is
\begin{equation}
  L_{\rm disk}=2\int_{r_{\rm ISCO}}^{\infty}2\pi r\,F(r)\,dr
  =\eta\,\dot M,
  \label{eq:disk-luminosity}
\end{equation}
with $\eta$ given in Eq.~\eqref{eq:binding-efficiency}. Relative to Schwarzschild,
\begin{equation}
  \frac{L_{\rm disk}}{L_{\rm Schw}}=\frac{\eta}{\eta_{\rm Schw}}
  \qquad (\text{for fixed }\dot M).
\end{equation}

A useful inverse relation for data interpretation is
\begin{equation}
  \dot M=\frac{L_{\rm disk}}{\eta}.
\end{equation}
Hence, if an observed source luminosity is fixed, models with larger deformation parameter $\beta$ require a smaller mass inflow rate than Schwarzschild, while larger $\nu$ reduces this difference. This gives a clear accretion-based channel---complementary to shadow and ringdown constraints---to infer how close the metric is to the extremal deformation bound.

To make the parameter dependence explicit, define the reduced flux profile
\begin{equation}
  \tilde F(x)\equiv\frac{16\pi M^2}{\dot M}F(r)=\mathcal{F}(x;\mu,\nu,\beta),
\end{equation}
and denote by $x_{\rm peak}$ the radius where $\tilde F$ is maximal. Representative numerical values are listed below: See Table~\ref{tab:accretion-beta-data} for fixed $(\mu,\nu)=(3,2)$ with varying deformation and Table~\ref{tab:accretion-nu-data} for fixed $(\mu,\beta)=(3,0.10)$ with varying interpolation exponent.

The previous ISCO analysis is directly connected to accretion observables:
\begin{itemize}
  \item Increasing $\beta$ (toward extremality) increases $\eta$, so for fixed $\dot M$ the total radiative output is enhanced and the inner disk spectrum becomes harder (higher peak $T_{\rm eff}$).
  \item Increasing $\nu$ suppresses deformation effects, increases $r_{\rm ISCO}$, and drives both luminosity and spectral shape back toward Schwarzschild expectations.
  \item Increasing $\mu$ at fixed $(\nu,\beta)$ follows the same direction as the ISCO trends above: larger efficiency and stronger concentration of dissipation near the inner edge.
\end{itemize}

\begin{table}[t]
  \centering
  \caption{Illustrative accretion data for fixed $(\mu,\nu)=(3,2)$ and varying deformation $\beta$. The luminosity ratio is for fixed $\dot M$, so $L_{\rm disk}/L_{\rm Schw}=\eta/\eta_{\rm Schw}$.}
  \label{tab:accretion-beta-data}
  \begin{tabular}{c c c c c c}
    \hline
    $\beta$ & $x_{\rm ISCO}$ & $\eta$ & $L_{\rm disk}/L_{\rm Schw}$ & $x_{\rm peak}$ & $\tilde F_{\rm max}$ \\
    \hline
    0.00 & 3.000 & 0.05719 & 1.000 & 4.775 & $6.88\times10^{-4}$ \\
    0.10 & 2.968 & 0.05773 & 1.009 & 4.725 & $7.08\times10^{-4}$ \\
    0.20 & 2.867 & 0.05946 & 1.040 & 4.570 & $7.80\times10^{-4}$ \\
    0.30 & 2.679 & 0.06296 & 1.101 & 4.278 & $9.39\times10^{-4}$ \\
    0.36 & 2.504 & 0.06653 & 1.163 & 4.009 & $1.12\times10^{-3}$ \\
    \hline
  \end{tabular}
\end{table}

\begin{table}[t]
  \centering
  \caption{Illustrative accretion data for fixed $(\mu,\beta)=(3,0.10)$ and varying interpolation exponent $\nu$. Increasing $\nu$ drives the model back toward Schwarzschild-like flux and efficiency.}
  \label{tab:accretion-nu-data}
  \begin{tabular}{c c c c c c}
    \hline
    $\nu$ & $x_{\rm ISCO}$ & $\eta$ & $L_{\rm disk}/L_{\rm Schw}$ & $x_{\rm peak}$ & $\tilde F_{\rm max}$ \\
    \hline
    1 & 2.002 & 0.08203 & 1.434 & 3.199 & $2.17\times10^{-3}$ \\
    2 & 2.968 & 0.05773 & 1.009 & 4.725 & $7.08\times10^{-4}$ \\
    3 & 2.999 & 0.05721 & 1.000 & 4.773 & $6.88\times10^{-4}$ \\
    4 & 3.000 & 0.05719 & 1.000 & 4.774 & $6.88\times10^{-4}$ \\
    6 & 3.000 & 0.05719 & 1.000 & 4.775 & $6.88\times10^{-4}$ \\
    \hline
  \end{tabular}
\end{table}

These datasets quantify the analytical trends discussed above: stronger deformation increases both the peak local dissipation and total radiative efficiency, while larger $\nu$ suppresses these deviations and rapidly restores the Schwarzschild benchmark.

\section{Conclusions}

Here we analyze various characteristics of regular black holes in non-polynomial quasi-topological gravity, including the Hawking temperature, Lyapunov exponents, shadow radii, binding energy, and accretion-related parameters. These quantities probe the spacetime geometry from the event horizon up to the radius of the innermost stable circular orbit, providing insight into how observable properties depend on the parameters of both the theory and the black hole.

The main conclusions are:
\begin{itemize}
  \item The physically relevant black-hole sector is sharply delimited by $\nu>0$, $\mu\ge3$, and $0<\beta\le\beta_c$, with $\beta_c$ setting the extremal boundary.
  \item The Hawking temperature decreases with stronger deformation and vanishes at extremality, while increasing $\nu$ restores Schwarzschild thermodynamics.
  \item The shadow radius follows the same qualitative trend as temperature: stronger deformation shrinks the shadow, whereas larger $\nu$ drives it back to the Schwarzschild limit.
  \item The photon-sphere Lyapunov exponent is likewise suppressed by deformation and approaches the Schwarzschild benchmark as $\nu$ increases.
  \item The ISCO binding efficiency behaves oppositely, increasing with deformation and therefore indicating potentially stronger accretion power in the deformed regime.
  \item In the Novikov--Thorne thin-disk model, this efficiency trend propagates directly to observables: larger deformation enhances disk luminosity and hardens the inner thermal profile at fixed $\dot M$, whereas larger $\nu$ restores Schwarzschild-like accretion signatures.
  \item Quantitatively, for $(\mu,\nu)=(3,2)$ and $\beta:0\to0.36$, the efficiency increases $\eta:0.05719\to0.06653$, the luminosity ratio rises $L_{\rm disk}/L_{\rm Schw}:1.000\to1.163$, the flux peak shifts inward $x_{\rm peak}:4.775\to4.009$, and the reduced peak flux grows $\tilde F_{\rm max}:6.88\times10^{-4}\to1.12\times10^{-3}$.
  \item For fixed $(\mu,\beta)=(3,0.10)$, increasing $\nu$ rapidly restores Schwarzschild-like accretion behavior: $\eta$ and $\tilde F_{\rm max}$ approach their Schwarzschild values by $\nu\gtrsim3$.
\end{itemize}

Taken together, these observables define a consistent phenomenological pattern: $\beta$ controls how far the geometry departs from Schwarzschild behavior, while $\nu$ controls how rapidly that departure is diluted. The added flux and luminosity datasets make this pattern directly testable in accretion modeling, providing a practical framework for confronting regular black-hole metrics with combined thermodynamic, optical, dynamical, and disk-emission constraints.

\bibliographystyle{unsrtnat}
\bibliography{references}

\end{document}